\documentclass[a4paper]{jpconf}
\usepackage{cite}
\usepackage{amsmath}
\begin{document}
\title{Ternary generalization of Heisenberg's algebra}

\author{Richard Kerner}

\address{Laboratoire de Physique Th\'eorique de la Mati\`ere Condens\'ee (LPTMC), Univesit\'e Pierre et Marie Curie - CNRS UMR 7600
Tour 23-13, 5-\`eme \'etage, Bo\^{i}te Courrier 121, 4 Place Jussieu, 75005 Paris, FRANCE}

\ead{richard.kerner@upmc.fr}
\vskip 0.7cm

{\it Dedicated to Bogdan Mielnik for the first 50 years of his scientific career}
\vskip 0.4cm

\begin{abstract}

A concise study of ternary and cubic algebras with $Z_3$ grading is presented.
We discuss some underlying ideas leading to the conclusion that the discrete symmetry
group of permutations of three objects, $S_3$, and its abelian subgroup $Z_3$ may
play an important role in quantum physics. We show then how most of important algebras
with $Z_2$ grading can be generalized with ternary composition laws combined
with a $Z_3$ grading.

We investigate in particular a ternary, $Z_3$-graded generalization of
the Heisenberg algebra. It turns out that introducing a non-trivial cubic root
of unity, $j = e^{\frac{2 \pi i}{3}}$, one can define two types of creation operators
instead of one, accompanying the usual annihilation operator. The two creation operators are
non-hermitian, but they are mutually conjugate. Together, the three operators form a ternary algebra, 
and some of their cubic combinations generate the usual Heisenberg algebra.

An analogue of Hamiltonian operator is constructed by analogy with the usual harmonic oscillator,
and some properties of its eigenfunctions are briefly discussed.

\end{abstract}

\section{Introduction}

More than fifty years ago Bogdan Mielnik has published several fundamental papers devoted to the analysis
of physical and mathematical bases of quantum mechanics. The titles of these articles speak for themselves:
the first two were named {\it ``Geometry of quantum states''} (\cite{Mielnik1968}) and {\it ``Theory of Filters''}
\cite{Mielnik1969}, and the third one was {\it ``Generalized Quantum Mechanics''} \cite{Mielnik1974}.

In these articles, remarkable for the depth and clarity of developed argumentation as well as for their mathematical
rigor, Bogdan Mielnik followed Bohr's interpretation of quantum mechanics rather than Einstein's point of view.
Both approaches displayed in their famous discussions \cite{Bohr1927,Lurcat,Lande}. The essence
of Bohr's approach was that neither a quantum object nor the classical object serving as measuring device can be
totally separated from each other; any measurement describes the {\it interaction} between the quantum object and the
observer. Einstein, on the other hand, in his opposition to Bohr, advocated a more classical idea of reality of
quantum object independent of observer. In order to refute Bohr's arguments, Einstein (together with his
collaborators, Boris Podolsky and Nathan Rosen) proposed an ideal experiment that would lead to a paradox
if Bohr's interpretation of quantum mechanics were right. The paradox consisted in the possibility of a measurement
performed on an atomic system to determine the issue of another measurement even in case of a space-like separation
excluding any transfer of information. For example, in spite of the apparent randomness of photon emission by a single
atom, according to Bohr, if a first photon emitted was found in a state with given polarization, the next photon
will be found with a polarization perpendicular to the first one, as if the system ``knew'' in advance what to do
next. Einstein thought that if such were the case, this would reveal the existence of {\it hidden parameters}
if the following three principles were conserved: {\it 1.~Causality, 2.~Relativity} (no physical influence can travel
faster than the speed of light $c$) 3. {\it Locality} (the absence of  action at a distance).

   In 1964 John Bell \cite{Bell1964,Bell1966} formulated a theorem which gives a possibility to check the
simultaneous validity of Einstein's point of view. The {\it Bell inequalities} introduced in these papers provide a criterion
discriminating between quantum theory with hidden parameters, advocated by Einstein, and the probabilistic interpretation
defended by Bohr. It took some time (more than twenty years) before a series of experiments carried out by A. Aspect and
collaborators \cite{Aspect} has proven the existence of the {\it entangled states} of photons, thus confirming the non-local
character of quantum physics, at least in some of its manifestations.

All that occurred long time after Bogdan Mielnik's first published papers; nevertheless he paid a great attention
to mathematical and physical postulates of quantum theory and analyzed with an extremely acute vision the
intricate framework of this fundamental branch of knowledge. Although he was active in other domains of physics, including 
General Relativity, his main and most original contributions concerned quantum theory and its possible extentions,
including quaternionic and non-associative structures. 

The models we shall display here represent a similar exploration effort, partly inspired by Bogdan Mielnik's work 
\cite{Mielnik1974}, with a novel algebraic approach based on {\it ternary} and {$Z_3-graded$} algebras. 
Before we proceed farther, let us expose the motivation for studying {\it ternary} and $Z_3$-{\it graded} generalizations
of Heisenberg's algebra. 

Quantum Mechanics provided an unexpected backing to ancient idea expressed by Plato in the seventh book of ``Republic''
known under the name of ``Allegory of the Cave'' \cite{Plato,Cohen,Hall}. Under closer scrutiny, it
appears that continuous physical quantities of classical physics, described exclusively by real numbers, are the
result of the averaging of measurements of quantum nature, described by complex probability amplitudes. Which leads
to the conclusion that the underlying reality is of quantum and discrete nature, while the classical picture is due
to the characteristic properties of our perception. 

As a paradoxical conclusion, one is tempted to say that
the so-called {\it complex numbers} deserve to be called ``real'', because they reflect the deepest  properties
of quantum physics, while the {\it real numbers} with which we describe our perceptible world appear to be the result
of averaging over great number of microscopic events, and therefore deserve better the adjective ``synthetic'' or ``imaginary''.

Thus, the underlying physical realm is described in terms of states which are vectors, or rather {\it rays}, in a Hilbert space over complex numbers.
One of the most striking properties of this space of states is the presence of universal discrete symmetry due to the representations
of the simplest permutation group $Z_2$. All elementary particles we know belong to two distinct classes: fermions or bosons. The first
ones ensure the stability of matter due to the Pauli exclusion principle \cite{Pauli1}, according to which two electrons cannot be in the same state
characterized by identical quantum numbers.  This principle not only explains the structure of atoms and
therefore the entire content of the periodic table of elements, but it also guarantees 
their stability by preventing collapse, as suggested by Ehrenfest~\cite{Dyson1},
and proved later by Dyson~\cite{Dyson2}. The relationship between the
exclusion principle and particle's spin, known under the name of the ``spin-and-statistic theorem'',
represents one of the deepest results in quantum field theory.

In purely algebraic terms Pauli's exclusion principle amounts to the anti-symmetry of 
wave functions describing two coexisting particle states. The easiest way to see how the principle
works is to apply Dirac's formalism in which wave functions of particles in given state are
obtained as products between the ``bra'' and ``ket'' vectors.


The wave function of a two-particle state of which one is in the state $\mid 1>$ and another
in the state $\mid 2 >$ (all other observables supposed to be the same for both states) is represented by a superposition
\begin{equation}
\mid \psi > =  \Phi (1,2) \, ( \mid 1> \otimes \mid 2>).
\label{xy-state}
\end{equation}
It is clear that if the wave function $\Phi (1,2)$ is anti-symmetric, i.e. if it satisfies $\Phi (1,2) = - \Phi (2,1),$
then $\Phi (1,1) = 0$ and such states have vanishing both their wave function and probability.
It is easy to prove using the superposition principle, that this condition is not only
sufficient, but also necessary.

After second quantization, when the states are obtained with creation and annihilation operators acting
on the vacuum, the anti-symmetry is encoded in the anti-commutation relations
\begin{equation}
a^{\dagger} (1) a^{\dagger} (2) + a^{\dagger} (2) a^{\dagger} (1) = 0, \; \; a (1) a(2) + a (2) a (1) = 0
\label{anticommutation}
\end{equation}
The bottom line is that the Hilbert space of fermionic states is always divided in two sectors corresponding to the
anti-commutation of creation of dichotomic spin state, admitting only two values which are labeled $+\frac{1}{2}$ and
$-\frac{1}{2}$. The anti-commuting character of their operator algebra of observables is represented by the antisymmetric
tensor $\epsilon_{\alpha \beta} = - \epsilon_{\beta \alpha}, \; \; \; \alpha, \beta = 1,2$. The exclusion principle being universal, 
it is natural to require that it should be independent of the choice of a basis in the Hilbert space of states. 
Therefore, if the states undergo a linear transformation 
\begin{equation}
\mid \psi^{\alpha}> \rightarrow \mid \psi_{\beta'}> = S^{\alpha}_{\beta'} \, \mid \psi_{\alpha}>,
\label{Stransform1}
\end{equation}
the anti-symmetric form encoding the exclusion principle should remain the same as before:
\begin{equation}  
 \epsilon_{\alpha' \beta'} = S^{\alpha}_{\alpha'} S^{\beta}_{\beta'} \epsilon_{\alpha \beta}, \; \; \; \text{ with } \; \; \; 
\epsilon_{1' 2'} = - \epsilon_{2' 1'} = 1, \; \; \; \epsilon_{1' 1' } = 0, \; \; \epsilon_{2' 2'} = 0.
\label{epsinv}
\end{equation}
This invariance condition, akin to the invariance of the metric tensor $\eta_{\mu \nu}$ of the Minkowskian
spacetime, defines the invariance group. It is easy to see that in this case the  effect of
(\ref{epsinv})  leads to the definition of the $SL(2, {\bf C})$ group. It is enough to
check one of the four equations (\ref{epsinv}), e.g. choosing $\alpha' = 1, \; \beta' = 2$. We get then
\begin{equation}
\epsilon_{1' 2'} = 1 = S^{\alpha}_{1'} S^{\beta}_{2'} \; \epsilon_{\alpha \beta} =
S^{1}_{1'} S^{2}_{2'} - S^{2}_{1'} S^{1}_{2'} = {\rm det} S = 1.
\label{defS1}
\end{equation}
The other three choices of index values in (\ref{epsinv}) are either redundant, or trivial, i.e. leading
to the identity $0 = 0$. 
The conjugate matrices span an inequivalent representation of the $SL(2, {\bf C})$ group, labeled by dotted indeces;
the antisymmetric $2$-form $\epsilon_{{\dot{\alpha}} {\dot{\beta}}}$ leads to the same result when its invariance is required:
$\epsilon_{{\dot 1} {\dot 2}} = - \epsilon_{{\dot 2} {\dot 1}} = 1, \; \;  \epsilon_{{\dot 1} {\dot 1}} = 0, 
\; \; \epsilon_{{\dot 2} {\dot 2}} = 0; \; \; \; 
\epsilon_{{\dot 1}' {\dot 2}'} =  {\bar{S}}^{{\dot{\alpha}}}_{{\dot{\alpha}}'} {\bar{S}}^{{\dot{\beta}}}_{{\dot{2}}'} 
\; \epsilon_{{\dot{\alpha}}{\dot{\beta}}}$
Spinors transforming with the $SL(2,{\bf C})$ group do not represent observable physical quantities; measurable quantities are formed
by their quadratic and hermitian combinations, like e.g. a four-vector ${\psi}^{\dagger} \; \gamma^{\mu} \psi $, transforming under 
the vector representation of the Lorentz group. 
This can suggest that the very origin of the Lorentz transformations resides in the discrete properties of elementary particles,
and not vice versa; more precisely, $SL(2, {\bf C})$ appears as the invariance group of the non-trivial action of the $Z_2$-group.
 
The mysterious properties of quarks, observable only by packs of three, or by quark-antiquark pairs, call for a non-trivial application of the $Z_3$-group,
or even the entire $S_3$ group. Mathematical constructions acknowledging these properties via introduction of $Z_3$-graded  algebras have
been proposed in~\cite{Kerner1991,Kerner1992,Abramov1997} and developed further in~\cite{Kerner2001,RKOS}.
A third-order generalization of Dirac equation was proposed, with waves that could propagate only as tensor products of three elementary
solutions with complex wave vectors each.

In what follows, we shall present a $Z_3$-graded analogue of Heisenberg's algebra and corresponding generalization of quantum oscillator.

\section{Cubic and ternary $Z_3$ graded algebras}

Our goal being a ternary generalization of Heisenberg's algebra, let us start with recalling basic facts about ternary algebras 
\cite{Gelfand,VainKer}
The usual definition of an algebra involves a linear space ${\cal{A}}$ (over real or complex numbers) endowed with a {\it binary} constitutive relations
\begin{equation}
{\cal{A}} \times {\cal{A}} \rightarrow {\cal{A}}
\label{algebradef}
\end{equation}
In a finite dimensional case, dim ${\cal{A}}$ = N, in a chosen basis ${\bf e}_1, {\bf e}_2, ..., {\bf e}_N$, the constitutive relations (\ref{algebradef})
can be encoded in {\it structure constants} $f^k_{ij}$ as follows:
\begin{equation}
{\bf e}_i  {\bf e}_j = f^k_{ij} \, {\bf e}_k.
\label{structure2}
\end{equation}
With the help of these structure constants all essential properties of a given algebra can be expressed, e.g. they will define a {\it Lie algebra} if they
are antisymmetric and satisfy the Jacobi identity:
\begin{equation}
f^k_{ij} = - f^k_{ji}, \; \; \; \; f^k_{im} f^m_{jl} + f^k_{jm} f^m_{li} + f^k_{lm} f^m_{ij} = 0,
\label{structureLie}
\end{equation}
whereas an abelian algebra will have its structure constants symmetric, $f^k_{ij} = f^k_{ji}$.

Usually, when we speak of algebras, we mean {\it binary algebras}, understanding that they are defined via {\it quadratic} 
constitutive relations (\ref{structure2}). On such algebras the notion of $Z_2-grading$ can be naturally introduced.
An algebra ${\cal{A}}$ is called a $Z_2-graded$  algebra if it is a direct sum of two parts, with symmetric (abelian) and
anti-symmetric product respectively, 
\begin{equation}
{\cal{A}} = {\cal{A}}_0 \oplus {\cal{A}}_1,
\label{Agraded}
\end{equation}
with {\it grade} of an element being $0$ if it belongs to ${\cal{A}}_0$, and $1$ if it belongs to ${\cal{A}}_1$.
Under the multiplication in a $Z_2$-graded algebra the grades add up reproducing the composition law of the $Z_2$
permutation group: if the grade of an element $A$ is $a$, and that of the element $B$ is $b$, then the grade of their
product will be $a+b$ {\it modulo}~2:
\begin{equation}
\operatorname{grade} (AB) = \operatorname{grade} (A) + \operatorname{grade} (B), \; \; \; \; \text{so that} \; \; \; AB = (-1)^{ab} BA.
\label{ABBAgraded}
\end{equation}
It is worthwhile to notice at this point that the above relationship can be written in an alternative form, with all
the expressions on the left hand side as follows:
\begin{equation}
A B - (-1)^{\alpha \beta} B A = 0, \; \; \text{ or } \; \; A B + (-1)^{(\alpha \beta + 1)} B A = 0
\label{comanticom5}
\end{equation}
The equivalence between these two alternative definitions of commutation (anticommutation) relations inside a $Z_2$-graded algebra
is no more possible if by analogy we want to impose {\it cubic} relations on algebras with $Z_3$-symmetry properties, in which
the non-trivial cubic root of unity, $j = e^{\frac{2 \pi i}{3}}$ plays the role similar to that of $-1$ in the binary relations
displaying a $Z_2$-symmetry. 

The $Z_3$ cyclic group is an abelian subgroup of the $S_3$ symmetry group of permutations of three objects. The $S_3$ groups contains {\it six } elements,
including the group unity $e$ (the identity permutation, leaving all objects in place: $(abc) \rightarrow (abc)$, the two cyclic permutations
$$ (abc) \rightarrow (bca) \; \; \; \text{ and } \; \; \; (abc) \rightarrow (cab),$$
and three odd permutations,
$$(abc) \rightarrow (cba), \; \; \; (abc) \rightarrow (bac) \; \; \; \text{ and } \; \; \; (abc) \rightarrow (acb).$$

There was a unique definition of {\it commutative} binary algebras given in two equivalent forms,
\begin{equation}
xy + (-1) yx = 0 \; \; \; \; \text{ or } \; \; \; \; \; xy = yx.
\label{comm2}
\end{equation}
In the case of cubic algebras \cite{VainKer} we have the following four generalizations of the notion of {\it commutative} algebras: 
\vskip 0.2cm
\indent
a) Generalizing the first form of the commutativity relation (\ref{comm2}), which amounts to replacing the $-1$ generator of $Z_2$ by
$j$-generator of $Z_3$ and binary products by products of three elements, we get
\begin{equation}
S: \; \; x^{\mu} x^{\nu} x^{\lambda} + j \;  x^{\nu} x^{\lambda} x^{\mu} + j^2 \; x^{\lambda} x^{\mu} x^{\nu} = 0,
\label{defS}
\end{equation}
where $j = e^{\frac{2 \pi i}{3}}$ is a primitive third root of unity.
\vskip 0.2cm
\indent
b) Another primitive third root, $j^2 = e^{\frac{4 \pi i}{3}}$ can be used in place of the former one; this will define the conjugate algebra
${\bar{S}}$, satisfying the following cubic constitutive relations:
\begin{equation}
{\bar S} : \; \; x^{\mu} x^{\nu} x^{\lambda} + j^2 \;  x^{\nu} x^{\lambda} x^{\mu} + j \; x^{\lambda} x^{\mu} x^{\nu} = 0.
\label{defSbar}
\end{equation}
Clearly enough, both algebras are infinitely-dimensional and have the same structure. Each of them is a possible generalization of infinitely-dimensional
algebra of usual commuting variables with a finite number of generators. In the usual $Z_2$-graded case such algebras are just polynomials in variables
$x^1, \; x^2, ... x^N$; the algebras $S$ and ${\bar{S}}$ defined above are also spanned by polynomials, but with different symmetry properties, and
 as a consequence, with different dimensions corresponding to a given power. 
\vskip 0.2cm
\indent
c) Then we can impose the following ``weak'' commutation, valid only for cyclic permutations of factors:
\begin{equation}
S_1 : \; \; x^{\mu} x^{\nu} x^{\lambda} = x^{\nu} x^{\lambda} x^{\mu} \neq x^{\nu} x^{\mu} x^{\lambda},
\end{equation} 
\vskip 0.2cm
\indent
d) Finally, we can impose the following ``strong'' commutation, valid for arbitrary (even or odd) permutations of three factors:
\begin{equation}
S_0: \; \; \; x^{\mu} x^{\nu} x^{\lambda} = x^{\nu} x^{\lambda} x^{\mu} = x^{\nu} x^{\mu} x^{\lambda}
\end{equation}
The four different associative algebras with cubic commutation relations can be represented in the following diagram, in which all arrows correspond
to {\it surjective homomorphisms}. The commuting generators can be given the common grade $0$.
\vskip 0.2cm
\centerline{$ S \hskip 1.5cm {\bar{S}}$}
\centerline{$ \searrow \hskip 0.5cm \swarrow$}
\centerline{$S_1$}
\centerline{$\downarrow$}
\centerline{$S_0$}
\vskip 0.2cm
Let us turn now to the $Z_3$ generalization of anti-commuting generators, which in the usual homogeneous case with $Z_2$-grading define Grassmann algebras.
Here, too, we have four different choices:   
\vskip 0.2cm
\indent
a) The ``strong'' cubic anti-commutation,
\begin{equation}
\Lambda_0 : \; \; \; {\displaystyle{\Sigma_{\pi \in S_3} \; \theta^{\pi (A)} \theta^{\pi (B)} \theta^{\pi (C)}}} = 0,
\end{equation}
i.e., the sum of {\it all} permutations of three factors, even and odd ones, must vanish.
\vskip 0.2cm
\indent
b) The somewhat weaker ``cyclic'' anti-commutation relation,
\begin{equation}
\Lambda_1 : \; \; \; \theta^A \theta^B \theta^C + \theta^B \theta^C \theta^A + \theta^C \theta^A \theta^B = 0,
\end{equation}
i.e., the sum of {\it cyclic} permutations of three elements must vanish. The same independent relation for the odd combination $\theta^C \theta^B \theta^A$
holds separately.
\vskip 0.2cm
\indent
c) The $j$-skew-symmetric algebra:
\begin{equation}
\Lambda : \theta^A \theta^B \theta^C = j \; \theta^B \theta^C \theta^A.
\end{equation}
and its conjugate algebra ${\bar{\Lambda}}$, isomorphic with $\Lambda$, which we distinguish by putting a bar on the generators and using dotted indices:
\vskip 0.2cm
\indent
d) The $j^2$-skew-symmetric algebra:
\begin{equation}
{\bar{\Lambda}} : \; \; \; {\bar{\theta}}^{\dot{A}} {\bar{\theta}}^{\dot{B}} {\bar{\theta}}^{\dot{C}} = 
j^2 {\bar{\theta}}^{\dot{B}} {\bar{\theta}}^{\dot{C}} {\bar{\theta}}^{\dot{A}}
\end{equation}
Both these algebras are finite dimensional. For $j$ or $j^2$-skew-symmetric algebras
with~$N$ generators the dimensions of their subspaces of given polynomial order are given by the following generating function:
\begin{equation}
H(t) = 1 + N t + N^2 t^2 + \frac{N (N-1)(N+1)}{3},
\end{equation}
where we include pure numbers (dimension $1$), the $N$ generators $\theta^A$ (or ${\bar{\theta}}^{\dot{B}}$), the $N^2$ independent quadratic combinations
$\theta^A \theta^B$ and $N(N-1)(N+1)/3$ products of three generators $\theta^A \theta^B \theta^C$.


The above four cubic generalization of Grassmann algebra are represented in the following diagram, in which all the arrows are surjective homomorphisms. 
\vskip 0.2cm
\centerline{$\Lambda_0$}
\centerline{$\downarrow$}
\centerline{$\Lambda_1$}
\centerline{$ \swarrow \hskip 0.4cm \searrow$}
\centerline{$ \Lambda \hskip 1.5cm {\bar{\Lambda}}$}
\vskip 0.2cm
\section{Examples of $Z_3$-graded ternary algebras}

\subsection{The $Z_3$-graded analogue of Grassman algebra}

Let us introduce $N$ generators spanning a linear space over complex numbers,
satisfying the following cubic relations \cite{Kerner1991,Kerner1992}:

\begin{equation}
\theta^A \theta^B \theta^C = j \, \theta^B \theta^C \theta^A = j^2 \, \theta^C \theta^A \theta^B,
\label{ternary1}
\end{equation}
with $j = e^{2 i \pi/3}$, the primitive root of $1$. We have $1+j+j^2 = 0$ \; \; and ${\bar{j}} = j^2$.

Let us denote the algebra spanned by the  $\theta^A$ generators by ${\bf{\cal{A}}}$~\cite{Kerner1991,Kerner1992}.

We shall also introduce a similar set of {\it conjugate} generators, 
 ${\bar{\theta}}^{\dot{A}}$,
$\dot{A}, \dot{B},... = 1,2,...,N$, satisfying similar condition with $j^2$ replacing $j$:

\begin{equation}
{\bar{\theta}}^{\dot{A}} {\bar{\theta}}^{\dot{B}} {\bar{\theta}}^{\dot{C}} = 
j^2 \, {\bar{\theta}}^{\dot{B}} {\bar{\theta}}^{\dot{C}} {\bar{\theta}}^{\dot{A}} 
= j \, {\bar{\theta}}^{\dot{C}} {\bar{\theta}}^{\dot{A}} {\bar{\theta}}^{\dot{B}},
\label{ternary2}
\end{equation}

Let us denote this algebra by  ${\bar{\cal{A}}}$.

We shall endow the algebra ${\cal{A}} \oplus {\bar{\cal{A}}}$ it with a natural  $Z_3$ grading, considering the generators~$\theta^A$
as grade $1$ elements, their conjugates ${\bar{\theta}}^{\dot{A}}$ being of grade $2$.

The grades add up modulo $3$, so that the products 
 $\theta^{A} \theta^{B}$ span a linear
subspace of grade~$2$, and the cubic products 
$ \theta^A \theta^B \theta^C$ being of grade $0$.
Similarly, all quadratic expressions in conjugate generators, 
${\bar{\theta}}^{\dot{A}} {\bar{\theta}}^{\dot{B}}$
are of  grade $2 + 2 = 4_{\operatorname{mod}3} = 1$, whereas their cubic products are again of grade $0$, 
like the cubic products od $\theta^A$'s. \cite{Kerner2001}

Combined with the associativity, these cubic relations impose finite dimension on the
algebra generated by the $Z_3$ graded generators. As a matter of fact, cubic expressions are the
highest order that does not vanish identically. The proof is immediate: 
\begin{equation}
\theta^A \theta^B \theta^C \theta^D = j \, \theta^B \theta^C \theta^A \theta^D =
j^2 \, \theta^B \theta^A \theta^D \theta^C 
= j^3 \, \theta^A \theta^D \theta^B \theta^C =
j^4 \, \theta^A \theta^B \theta^C \theta^D, 
\label{quartic1}
\end{equation}
and because
$j^4 = j \neq 1$, the only solution is $\theta^A \theta^B \theta^C \theta^D = 0.$

\subsection{The $Z_3$ graded differential forms}

Instead of the usual exterior differential operator satisfying 
$d^2 = 0 $ , we can postulate its $Z_3$-graded generalization satisfying 
$$d^2 \neq 0, \; \; \;   d^3 f = 0 $$
The first differential of a smooth function $f (x^i)$ is as usual
$$df = \partial_i f \, dx^i, $$
 whereas the second differential is formally
$$d^2 f = ( \partial_k \partial_i f ) \; dx^k dx^i + (\partial_i f ) \; d^2 x^i $$

 We shall attribute the grade $1$ to the $1$-forms
$d x^i, \; (i,j,k = 1,2,...N)$ , and  grade $2$  to the forms
$d^2 x^i, \; (i,j,k = 1,2,...N) $ ; under associative multiplication of these forms the grades 
add up \textit{modulo}~$3$
$$\operatorname{grade} ( \omega   \; \theta) = \operatorname{grade} (\omega) + \operatorname{grade} (\theta) \; (\textit{modulo}~3).$$
 The $Z_3$-graded differential operator $d$ 
has the following property, compatible with grading we have chosen: 
\begin{gather*}
d (\omega \; \theta) = ( d \omega ) \, \theta + j^{\operatorname{grade}_{\omega}} \, \omega \, d \theta.\\
d^2 f = (\partial_i \partial_k f) d x^i d x^k + (\partial_i f ) \, d^2 x^i,\\
d^3 f = (\partial_m \partial_i \partial_k f ) d x^m d x^i d x^k +
( \partial_i \partial_k f ) d^2 x^i d x^k\\
+ j \; ( \partial_i \partial_k f ) d x^i d^2 x^k +
(\partial_k \partial_i f) d x^k d^2 x^i + (\partial_i f ) \, d^3 x^i .
\intertext{equivalent with}
d^3 f = (\partial_m \partial_i \partial_k f ) d x^m d x^i d x^k + ( \partial_i \partial_k f) [ d^2 x^k d x^i - j^2 \, d x^i d^2 x^k ] 
+ (\partial_i f ) \, d^3 x^i.
\end{gather*}

 Consequently, assuming that $d^3 x^k = 0$ and $d^3 f = 0$, to make 
the remaining terms vanish we must impose the following commutation relations 
on the products of forms:
\begin{gather*}
d x^i d x^k d x^m = j \, d x^k d x^m d x^i, \; \; \; \; \; 
d x^i d^2 x^k = j \, d^2 x^k d x^i,\\
\intertext{therefore}
d^2 x^k d x^i = j^2 \, d x^i d^2 x^k.
\end{gather*}

As in the case of the abstract $Z_3$-graded Grassmann algebra, the fourth order
expressions must vanish due to the associativity of the product: 
$$dx^i dx^k dx^l dx^m =0.$$
Consequently, we shall assume that also
$$d^2 x^i d^2 x^k = 0.$$
This completes the construction of algebra of $Z_3$-graded exterior forms.

\subsection{Ternary Clifford algebra}
\indent
Let us introduce the following three $3 \times 3$ matrices:
\begin{equation}
Q_1 = \begin{pmatrix} 0 & 1 & 0 \\ 0 & 0 & j \\ j^2 & 0 & 0 \end{pmatrix}, \; \; \; 
Q_2 = \begin{pmatrix} 0 & j & 0 \\ 0 & 0 & 1 \\ j^2 & 0 & 0 \end{pmatrix}, \; \; \; 
Q_3 = \begin{pmatrix} 0 & 1 & 0 \\ 0 & 0 & 1 \\ 1 & 0 & 0 
\end{pmatrix}, 
\label{threeQ}
\end{equation} 
and their hermitian conjugates
\begin{equation}
Q^{\dagger}_1 = \begin{pmatrix}  0 & 0 & j \\ 1 & 0 & 0 \\ 0 & j^2 & 0 \end{pmatrix}, \; \; \; 
Q^{\dagger}_2 = \begin{pmatrix} 0 & 0 & j \\ j^2 & 0 & 0 \\ 0 & 1 & 0 \end{pmatrix}, \; \; \; 
Q^{\dagger}_3 = \begin{pmatrix} 0 & 0 & 1 \\ 1 & 0 & 0 \\ 0 & 1 & 0 \end{pmatrix}. 
\label{threeQbar}
\end{equation} 
These matrices can be allowed natural $Z_3$ grading,
\begin{equation}
\operatorname{grade} (Q_k) = 1, \; \; \; \operatorname{grade} (Q^{\dagger}_k) = 2,
\label{gradeQ}
\end{equation}
The above matrices span a very interesting ternary algebra. Out of three independent $Z_3$-graded ternary
combinations, only one leads to a non-vanishing result. One can check without much effort that both $j$ and $j^2$ skew 
ternary commutators do vanish:
\begin{align*}
\{ Q_1, Q_2, Q_3 \}_j = Q_1 Q_2 Q_3 + j Q_2 Q_3 Q_1 + j^2 Q_3 Q_1 Q_2 &= 0,\\
\{ Q_1, Q_2, Q_3 \}_{j^2} = Q_1 Q_2 Q_3 + j^2 Q_2 Q_3 Q_1 + j Q_3 Q_1 Q_2 &= 0,
\end{align*}
and similarly for the odd permutation, $Q_2 Q_1 Q_3$.
On the contrary, the totally symmetric combination does not vanish; it is proportional to the $3 \times 3$ identity matrix ${\bf 1}$:
\begin{equation}
 Q_a Q_b Q_c + Q_b Q_c Q_a + Q_c Q_a Q_b = \eta_{abc} \, {\bf 1}, \; \; \; a,b,... = 1,2,3.
\label{anticom}
\end{equation}
with $\eta_{abc}$ given by the following non-zero components:
\begin{equation}
\eta_{111} = \eta_{222} = \eta_{333} = 1, \; \; \; \eta_{123} = \eta_{231} = \eta_{312} = j^2, \; \; \; 
\eta_{213} = \eta_{321} = \eta_{132} = j,
\label{defeta}
\end{equation}
all other components vanishing. The relation~(\ref{anticom}) may serve as the definition of \textit{ternary Clifford algebra}.

Another set of three matrices is formed by the hermitian conjugates of $Q_a$, which we shall endow with dotted indices ${\dot{a}}, {\dot{b}},...=1,2,3$:
$Q_{{\dot{a}}} = Q_a^{\dagger}$
satisfying conjugate identities
 \begin{equation}
 Q_{\dot{a}} Q_{\dot{b}} Q_{\dot{c}} + Q_{\dot{b}} Q_{\dot{c}} Q_{\dot{a}} + Q_{\dot{c}} Q_{\dot{a}} Q_{\dot{b}}
 = \eta_{{\dot{a}}{\dot{b}}{\dot{c}}} \, {\bf 1}, \; \; \; {\dot{a}}, {\dot{b}},... = 1,2,3.
\label{anticomdot}
\end{equation}
with $\eta_{{\dot{a}}{\dot{b}}{\dot{c}}} = {\bar{\eta}}_{cba}$.

It is obvious that any similarity transformation of the generators $Q_a$ will keep the ternary anti-commutator (\ref{defeta})
invariant. As a matter of fact, if we define ${\tilde{Q}}_b = P^{-1} Q_b P$, with $P$ a non-singular $3 \times 3$ matrix,
 the new set of generators will satisfy the same ternary relations, because 
$${\tilde{Q}}_a {\tilde{Q}}_b {\tilde{Q}}_c = P^{-1} Q_a P P^{-1} Q_b P P^{-1} Q_c P = P^{-1} (Q_a Q_b Q_c) P,$$
and on the right-hand side we have the unit matrix which commutes with all other matrices, so that $P^{-1} \; {\bf 1} \; P  = {\bf 1}$.

\subsection{Ternary $Z_3$-graded commutator}

 In any associative algebra ${\cal{A}}$ one can introduce a new binary operation,
the {\it commutator}, using the generator of the  $Z_2$ group in form of 
multiplication by $-1$ :
\begin{equation}
X, Y \in A \rightarrow [X, Y ] = X Y + (-1) Y X = X Y - Y X.
\label{Z2commdef}
\end{equation}
In the case of the  $Z_2$ group the generator of its representation
on complex numbers was equal to  $-1$; note that $ -1 + (-1)^2 = 0$
In the case of the  $Z_3$ group, the generator of its complex representation can be
chosen to be  $ j = e^{\frac{2 \pi i}{3}}$ , with $j + j^2 + j^3 = 0 $

Consider the following cubic combination defined on an associative algebra ${\cal{A}}$: 
$$ X, Y, Z \in {\cal{A}}, \; \; \; \; \; \; 
\{ X, Y, Z \} := XYZ + j YZX + j^2 ZXY.$$
One obviously has:
$$ \{ X, Y, Z \} = j \{ Y, Z, X \} = j^2 \{ Z, X, Y \}, \; \; \; 
\text{ and consequently } \; \; \; 
 \{ X, X, X \} = 0.$$
In the case when ${\cal{A}}$ is a  {\it unital} algebra, i.e. it contains a unit element  {\bf 1} such that 
${\bf 1} X = X {\bf 1} = X$ for any  $X \in {\cal{A}}$, a unique Lie algebra is naturally generated by the cubic commutator:
\begin{multline*}
\{ X, 1, Y \} = X \cdot 1 \cdot Y + j \; 1 \cdot Y \cdot X + j^2 \; Y \cdot X \cdot 1\\
= XY + j \; YX + j^2 \; YX = XY + (j+j^2) YX = XY - YX = [X,Y].
\end{multline*}
 The following simple example  of a $Z_3$ cubic algebra can be constructed with $2 \times 2$ complex matrices.
Consider the Lie algebra spanned by three Pauli's matrices:
$$
\sigma_1 = \begin{pmatrix} 0 & 1 \\ 1 & 0 \end{pmatrix}, \; \; \; 
\sigma_2 = \begin{pmatrix} 0 & -i \\ i & 0 \end{pmatrix}, \; \; \; 
\sigma_3 = \begin{pmatrix}1 & 0 \\ 0 & -1 \end{pmatrix} $$
satisfying the well known commutation relations:
$$[ \sigma_i , \sigma_j ] = C^k_{ij} \; \sigma_k, \; \; \; i,j,k = 1,2,3, \; \; \; C^k_{ij} = 2 i \epsilon_{kij}.$$
The enveloping algebra ${\cal{A}}_{\sigma}$  contains the unit matrix ${\bf  1}$:
$$\sigma_i \sigma_j = i \; \epsilon_{ijk} + \delta_{ij} {\bf 1} $$
Let us define the {\it cubic j-commutator} on the algebra ${\cal{A}}_{\sigma}$: 
$$\{ \sigma_i, \sigma_j, \sigma_k \} = \sigma_i \sigma_j \sigma_k + j \; \sigma_j \sigma_k \sigma_i 
+ j^2 \; \sigma_k \sigma_i \sigma_j, \;  \; j= e^{\frac{2 \pi i}{3}} .$$
This cubic algebra contains  three cubic subalgebras generated by two $\sigma$-matrices out of three: for example 
$$\{ \sigma_1, \sigma_2, \sigma_1 \} = - 2 \, \sigma_2, \; \; \; \{ \sigma_2 , \sigma_1 \, \sigma_2 \} = - 2 \, \sigma_1,$$
 and similarly for the couples $\sigma_2 , \sigma_3$   and $\sigma_3, \, \sigma_1$.
We have also $\{ \sigma_1, \, \sigma_2, \, \sigma_3 \} = 0.$

\section{Ternary Heisenberg algebra}
 \subsection{The $Z_2$ Heisenberg algebra}
Let us first remind the original construction of the Heisenberg algebra used in the analysis of {\it quantum harmonic oscillator}.
The Hamiltonian of classical harmonic oscillator, expressed in reduced dimensionless variables reads: 
$$H = \frac{p^2}{2} + \frac{x^2}{2}$$
after first quantization which attributes to canonical variables $(p, x)$ corresponding quantum operators acting
on the Hilbert space $L^2(R)$ of square-integrable functions
$$p \rightarrow {\hat{p}} = - i \hbar \frac{d}{dx}, \; \; \; \; x \rightarrow {\hat{x}}$$
The quantum version of the Hamiltonian is the hermitian operator ($\hbar=1$)
$${\hat{H}} = \frac{1}{2} \; \left[ - \frac{d^2}{d x^2} + x^2 \right] $$ 

The classical Heisenberg algebra~\cite{Schroedinger} is generated by the following two operators: 
$$a = \frac{1}{\sqrt{2}} \left( {\bar{x}} + i {\bar{p}} \right), \; \; \; \; a^{\dagger} = \frac{1}{\sqrt{2}} \left( {\bar{x}} - i {\bar{p}} \right),$$
(in terms of dimensionless operators ${\bar{x}}$ and ${\bar{p}}$ defined as follows:
$${\bar{x}} = \frac{x}{\lambda}, \; \; \; {\bar{p}} = \frac{\lambda \, p}{\hbar}.$$
with  $\lambda$ some unit of length) satisfying the well known commutation relations
$$ \left[ {\bar{x}}, \; {\bar{p}} \right] = i.
\; \; \; \;   [ a, \; a^{\dagger}] = 1.$$
From now on we follow original Schr\"odinger's presentation~\cite{Schroedinger}, without the factors $1/{\sqrt{2}}$ in $a$ and $a^{\dagger}$,
and with Hamiltonian without the factor $1/2$.
Now the quantum version of he Hamiltonian becomes a hermitian operator, which can be expressed by means of the
operators $a$ and $a^{\dagger}$: 
$${\hat{H}} = - \frac{d^2}{d x^2} + x^2 = a a^{\dagger} + a^{\dagger} a,$$ 
The eigenfunctions of this operator, corresponding to fixed values of energy $E$, are obtained by defining first the 
lowest energy state corresponding to the zero-eigenvalue state of the operator $a$, $a \mid 0 > = 0$.
In coordinate representation the zero state is represented by a Gaussian function:  
\begin{equation}
a \mid 0 > = 0 \; \; \rightarrow \bigg[\frac{d}{dx} + x \bigg] \, \Phi_0 \; (x) = 0 \; \; \; \rightarrow \Phi_0 \; (x) = e^{- \frac{x^2}{2}}.
\label{eigena}
\end{equation}
Positive eigenvalue functions of the Hamiltonian ${\hat{H}}$ are then obtained by acting on $f(x)$ with consecutive powers of $a^{\dagger}$~\cite{Schroedinger}.
These eigenfunctions span the well-known discrete energy spectrum (for bound states with negative total energy). The hermitian operators
${\hat{p}}$ and ${\hat{x}}$ could be replaced in this construction by non-hermitian combinations $a = ip + x, \; \; a^{\dagger} = - i p + x$
because their quadratic combination $H = a^{\dagger} a + a a^{\dagger}$ is hermitian and positive definite. Due to the commutation relations
between $a^{\dagger}$ and $a$, it can be easily checked that the new function obtained by the action of $a^{\dagger}$ on the vacuum,
which is 
\begin{equation}
a^{\dagger} \mid 0 > = \mid 1 > \; \; \rightarrow \; \;  \bigg[ - \frac{d}{dx} + x \bigg] \, e^{-\frac{x^2}{2}} = 2 x \, e^{- \frac{x^2}{2}} = \Phi_1,
\label{Phione}
\end{equation}
is an eigenfunction of the operator $H$:
\begin{equation}
H \, \mid \Phi_1 > = 3 \, \mid \Phi_1 > \; \; \text{ because } \; \; \bigg[ - \frac{d^2}{x^2} + x^2 \bigg] \, 2 x e^{-\frac{x^2}{2}} 
= 6 x \; e^{- \frac{x^2}{2}}.
\label{Eigenthree}
\end{equation}

\subsection{The $Z_3$ ternary Heisenberg algebra}

In the  $Z_2$-graded case the two independent
combinations of the operators  $\frac{d }{ d x}$ and  $x$
are given by the following non-singular matrix:
$$ \begin{pmatrix} 1 & 1 \\ -1 & 1 \end{pmatrix}.$$
 By analogy, in the  $Z_3$-graded case, the following
transformation matrix should be used, producing three independent combinations of the operators
$\frac{d }{ d x}, \; x$ and  ${\rm {\bf 1}}$: 
$$\begin{pmatrix} 1 & 1 & 1 \\ j & j^2 & 1 \\ j^2 & j & 1 \end{pmatrix} $$

A natural ternary generalization of Heisenberg's algebra should be spanned by three generators~\cite{Nambu} instead of two.
By analogy with the $Z_2$ case, the $Z_3$-generalization 
will be generated by the following  three operators: 
$$c_1 = \lambda \frac{d}{d x} + j x + j^2 {\mathrm 1}, \; \; \; \; \;
c_2 = \lambda \frac{d}{d x} + j^2 x + j {\mathrm 1}, \; \; \; \; \;
c_3 = \lambda \frac{d}{d x} + x +  {\mathrm 1},$$
 which span the following {\it binary} Lie algebra: 
$$ \left[ c_1, c_2 \right] = \lambda (j^2 - j) \; 1, \; \; \; \; \;
\left[ c_2, c_3 \right] = \lambda (1 - j^2) \; 1, \; \; \; \; \;
\left[ c_3, c_1 \right] = \lambda (j - 1 ) \; 1,$$

The following linear relations hold:
$$\frac{1}{3} \left( c_1 + c_2 + c_3 \right) = \lambda  \frac{d}{d x}, \; \; \; \; \;
\frac{1}{3} \left( j \; c_1 + j^2 \; c_2 + c_3 \right) = {\rm {\bf 1}}, \; \; \; \; \; 
\frac{1}{3} \left( j^2 \; c_1 + j c_2 + c_3 \right) = x,$$
 Here are the independent $3$-commutators between the generators of our algebra:
\begin{gather*}
\{ c_1, c_2, c_1 \} = - 3 \lambda \, c_1, \; \; \;  \{ c_2, c_1, c_2 \} =  3 \lambda \, c_2,\\
\{ c_2, c_3, c_2 \} = - 3 j \, \lambda \, c_2, \; \; \;  \{ c_3, c_2, c_3 \} =  3 j \lambda \, c_3,\\
\{ c_3, c_1, c_3 \} = - 3 j^2 \, \lambda \, c_3, \; \; \;  \{ c_1, c_3, c_1 \} =  3 j^2 \, \lambda \, c_1,\\
\{ c_2, c_3, c_1 \} = \lambda \, \left[ (1-j) c_1 + (j^2 -1) c_2 + (j - j^2) c_3 \right],\\
\{ c_1, c_3, c_2 \} = \lambda \, \left[ (j^2-j) c_1 + (j^2 -j) c_2 + (j^2 - j   ) c_3 \right],
\end{gather*}
If we add the unit operator  ${\rm {\bf 1}}$ to the three generators  $c_1, \; c_2$
and $c_3$, the ordinary commutators between the generators can be interpreted as ternary commutators involving the
unit operator $1$ in the middle: 
\begin{gather*}
[ c_1, c_2 ] = \{ c_1, 1, c_2 \} = \lambda (j^2 - j) \; 1,\\
[ c_2, c_3 ] = \{ c_2, 1, c_3 \} = \lambda (1 - j^2) \; 1,\\
[ c_3, c_1 ] = \{ c_3, 1 , c_1 \} = \lambda (j - 1 ) \; 1,\\
(c_1 c_3 c_2 + c_3 c_2 c_1 + c_2 c_1 c_3) - (c_2 c_3 c_1 + c_3 c_1 c_2 + c_1 c_2 c_3) = 3 \lambda (j - j^2).
\end{gather*}
after renormalization $c_k \rightarrow \frac{1}{\sqrt{3}} c_k$ and
setting $\lambda = -i$, the right-hand side of this cubic commutation relation will
become equal to $\hbar {\rm{\bf 1}}$. Normalized operators are as follows
 $$c_1 = \frac{1}{\sqrt{3}} \left[ \frac{ \sqrt{\hbar}}{i} \frac{d}{dx} + j x + j^2 1 \right], \; \; 
 c_2 = \frac{1}{\sqrt{3}} \left[ \frac{ \sqrt{\hbar}}{i} \frac{d}{dx} + j^2 x + j 1 \right], \; \; 
 c_3 = \frac{1}{\sqrt{3}} \left[ \frac{ \sqrt{\hbar}}{i} \frac{d}{dx} + x + 1 \right],$$

\section{Ternary analogue of quantum oscillator}

\subsection{Cubic combinations of the generators}

The $Z_3$-graded analogues of creation and annihilation operators  $a$ and $a^{\dagger}$ are the operators 
$c_3= c_3^{\dagger}, \; c_1$ and $c_2 = c_1^{\dagger}$. Imposing hermiticity means that the factor $\lambda$ in front
of the derivation must be pure imaginary, so we shall set $\lambda = -i$, as in the usual quantization scheme.

The expression $a a^{\dagger} + a^{\dagger} a $ can be interpreted as a sum of all $Z_2$ permutations, or all $S_2$
permutations, because here all permutations are cyclic, which is not the case of $Z_3$ and $S_3$ groups. The sum of
 all cyclic permutations of three operators $c_1 c_2 c_3$, or even all (six) permutations does not lead to any simple
expression. Moreover, in such combinations the complex representation of $Z_3$ does not appear.

However, the quadratic harmonic oscillator Hamiltonian $a a^{\dagger} + a^{\dagger} a $ can be represented in an
alternative manner, in which the representation of $Z_2$ by multiplications by $1$ and $-1$ does appear explicitly:
\begin{equation}
{\hat{H}} = \frac{1}{2} \; \left[ (a + a^{\dagger})^2 + (-1) (a - a^{\dagger})^2 \; \right] = a a^{\dagger} + a^{\dagger} a.
\label{Hquadnew}
\end{equation} 
Now the $Z_3$ generalization becomes obvious: we should form the sum of {\it cubes} of similar expressions with three
generators each, multiplied by $j$ and $j^2$ in all possible combinations so as to ensure the hermiticity of the resulting
expression, like in (\ref{Hquadnew}) above. Two expressions of this type can be formed with the generators $c_1, c_2$ and $c_3$,
\begin{align*}
&\frac{1}{27} \; [ (c_3 + c_1 + c_2)^3 + j (c_3 + j c_1 + j^2 c_2)^3 + j^2 (c_3 + j^2 c_1 + j c_2)^3 ],\\
\text{and }\;&\frac{1}{27} \; [ (c_3 + c_1 + c_2)^3 + j^2 (c_3 + j c_1 + j^2 c_2)^3 + j (c_3 + j^2 c_1 + j c_2)^3].
\end{align*}
Neither of these two combinations is hermitian, but their sum ${\hat{K}}_{Z_3}$ is (if we set $\lambda $ pure imaginary, e.g. $-i$):
\begin{equation}
{\hat{K}}_{Z_3} = 2 \lambda^3 \, \frac{d^3 }{dx^3} - x^3 - 1 =  2 i \frac{d^3 }{dx^3} - x^3 - 1.
\label{HatZ3}
\end{equation}
The expression $x^3 + 1$ appearing
 as the ``potential'' part factorizes as follows:
$$x^3 + 1 = (x + 1) (x +j) (x + j^2).$$
However, this operator does not seem to be a proper $Z_3$-graded generalization of quantum harmonic oscillator Hamiltonian. 
It is an odd function of momentum, and its physical dimension is a fractional power of energy, and its spectrum, although real,
is not bound from below because it is not positive-definite like the ordinary quadratic Hamiltonian.

The eigenvalue equation
$${\hat{K}}_{Z_3} f(x) = K f (x)$$
can be solved explicitly for the particular value of $K = -1$, which amounts to solve the following differential equation of third order:
\begin{equation}
2 i \frac{d^3 f}{d x^3} - x^3 \, f(x) = 0.
\label{eigenf}
\end{equation}
This equation can be solved by expanding $f(x)$ into a power series
$$f(x) = {\displaystyle{\sum_{k=0}^{\infty}}} \, c_k \, x^k.$$
The solution is given in terms of the {\it generalized hypergeometric function} $F([\hphantom{x}]; p,q; \xi)$.
In the convention used in ``Maple'' computing program, this symbol defines the following power series:
\begin{equation}
F([\hphantom{x}]; p,q; \xi) = {\displaystyle{\sum_{k=0}^{\infty}}} \, c_k x^k,\quad c_k = \frac{\Gamma(p) \Gamma(q)}{\Gamma(p+k) \Gamma(q+k)}.
\label{hyposum}
\end{equation}
The eigenfunction satisfying (\ref{eigenf}) is a linear combination of three independent solutions:
\begin{equation}
F ( [\hphantom{x}]; \frac{2}{3}, \frac{5}{6}; \frac{i x^6}{432} ), \; \; \; x \; F ( [\hphantom{x}]; \frac{5}{6}, \frac{7}{6}; \frac{i x^6}{432} ),
\; \; \; \text{ and } \; \; \;x^2 \;  F ( [\hphantom{x}]; \frac{7}{3}, \frac{4}{3}; \frac{i x^6}{432} ),
\label{threeF}
\end{equation}
corresponding to three different initial conditions, as it should be for a third order differential equation. All these eigenfunctions
are complex.

In the classical harmonic oscillator case the operators $a$ and $a^{\dagger}$ were not hermitian; nevertheless their combination
$a a^{\dagger} + a^{\dagger} a$ became hermitian. 
We could have as well abandoned the requirement of hermiticity, i.e. admit real value
for the factor $\lambda$ in the definition of our operators $c_k$, and still get real eigenvalues for the non-hermitian third order
operator, in the spirit of C.~Bender's work on the subject~\cite{Bender}. But the third-order differential operator is not a good
candidate for Hamiltonian of any kind, because it is not positive definite and its spectrum can not display a minimal value.
Moreover, in the operators $c_k$ the momentum operator ${\hat{p}} = - i d/dx$ can not be replaced by its real counterpart $d/dx$
without destroying the hermiticity. Such a substitution will be nevertheless possible if we introduce quadratic combination of
two third-order operators. 
This is why we propose to consider the following two third-order operators:
\begin{equation}
{\hat{K}}_1 = i p^3 + x^3 = - \frac{d^3}{d x^3} + x^3, \; \; \; \; {\hat{K}}_2 = -i  p^3 + x^3 =  \frac{d^3}{d x^3} + x^3.
\label{twoKs}
\end{equation} 
Then it is easy to see that one has
\begin{equation}
{\hat{K}}_1^{\dagger} = {\hat{K}}_2 \; \; \; \text{ and } \; \; \; \frac{1}{2} \left[ {\hat{K}}_1 {\hat{K}}_2 + {\hat{K}}_2 {\hat{K}}_1 \right]
 = {\hat{H}}_{Z_3} = \frac{d^6}{d x^6} + x^6.
\label{H6}
\end{equation}  
The zero-value eigenfunctions of the operator are of the same form as the eigenfunction (\ref{threeF}), but with real argument
$x^6/432$ instead of the imaginary one $i x^6/432$. 
The same operator can be represented as a cubic combination of second-order operators, one of which is the classical harmonic oscillator, 
the two remaining ones of the same form, but with complex conjugate potential parts:
\begin{equation}
{\hat{H_0}} = -\frac{d^2}{dx^2} + x^2, \; \; {\hat{H_1}} = -\frac{d^2}{dx^2} + j \;  x^2, \; \; {\hat{H_2}} = -\frac{d^2}{dx^2} + j^2 \; x^2, 
\label{ThreeH}
\end{equation} 
One easily checks that
\begin{equation}
{\hat{H_0}} {\hat{H_1}} {\hat{H_2}} + {\hat{H_1}} {\hat{H_2}} {\hat{H_0}} + {\hat{H_2}} {\hat{H_0}} {\hat{H_1}} = - \frac{d^6}{d x^6} + x^6.
\label{cyclicH}
\end{equation} 
The odd combination of the same kind gives the identical result.

This suggests that what we are describing here is a kind of third power of ordinary harmonic oscillator with its complexified
counterparts.  Each of the two ``exotic'' operators, ${\hat{H}}_1 = i {\hat{p}} + j x$ and ${\hat{H}}_2 = {\hat{H}}_1^{\dagger} = i {\hat{p}} + j^2 x$ 
displaying a complex spectrum, with the energies multiplied by $j$ and by $j^2$, respectively. This seems to indicate that the spectrum of the
cubic combination ${\hat{H}}_{Z_3}$ should be proportional to the products of the corresponding three eigenvalues, $E_0$, $E_1 = j E_0$
and $E_2 = j^2 E_0$, leading to the real eigenvalue $E_0 \cdot E_1 \cdot E_2 = j \cdot j^2 \cdot E_0^3 = E_0^3$.

We shall show now that the classical Bohr-Sommerfeld quantization scheme applied to the Hamiltonian ${\hat{H}}_{Z_3}$ confirms this hypothesis. 

\subsection{Bohr-Sommerfeld quantization}

In the case of a one-dimensional harmonic oscillator whose classical Hamiltonian is given by the quadratic expression
$$H (p, x) = \frac{p^2}{2 m} + \frac{k x^2}{2},$$ 
the Bohr-Sommerfeld quantization rule imposes the condition of finite discrete energy levels via the following rule:
\begin{equation}
\int_{\text{period}} \; p dx = n \hbar
\label{BSquant1}
\end{equation}
Expressing the momentum $p$ as a function of the coordinate $x$ by posing $H(p,x) = E =$ Constant,
 one gets the explicit expression of this integral:
\begin{equation}
\int_{\text{period}} \; p dx = 2  m \omega \; \int_{x_{min}}^{x_{max}} \; \sqrt{ \frac{2 E}{k} - x^2} \; dx = n \hbar,
\label{BSintexp}
\end{equation} 
where $\omega = \sqrt{k/m}$, and $E$ is the energy.
The integral is well known, and leads to the quantization rule
$E = \omega \hbar \left (n + \frac{1}{2} \right).$

Now let us apply a similar quantization condition to our ``cubic Hamiltonian''.
We start with the expression which is supposed to describe something that is an analogue of a Hamiltonian
 function. This expression is obtained from ternary generalization of Heisenberg's algebra:
\begin{equation}
H_{(3)} = \frac{p^6}{6 m^3} + \frac{k^3 x^6}{6}
\label{Ham3}
\end{equation}
The dimension of this expression is the third power of energy, i.e.,  $\operatorname{dim}[H] = [\text{energy}]^3$. 

We shall suppose that the expression (\ref{Ham3}) is constant in time. This gives the following
equation:
\begin{equation}
H_{(6)} = \frac{p^6}{6 m^3} + \frac{k^3 x^6}{6} = E^3 = \text{Const.}
\label{Ham3const}
\end{equation}
This constant enables us to get the following expression of $p$ as a function of $x$:
\begin{equation}
p = 6^{\frac{1}{6}} \, \sqrt{m k} \left[ \frac{6 E^3}{k^3} - x^6 \right]^{\frac{1}{6}} .
\label{pdex}
\end{equation}
It is easy to see that the corresponding trajectory is closed, and the motion must be periodic.

Now let us impose the Bohr-Sommerfeld quantization condition (\ref{BSquant1}), which is based on a quasi-classical approximation and gives very good agreement with observation
in the large quantum number limit, as shown e.g., in~\cite{Gorska2010},
 which in this case yields the explicit integral that can be evaluated with a little effort, yielding the following result
when we substitute for $p$ its expression (\ref{pdex}):
\begin{equation}
 \int_{\text{period}} p dx = 2 \, \int_{-x_{max}}^{+ x_{max}} p dx = 2 \sqrt{mk} \; \int_{-x_{max}}^{+ x_{max}} \, \left[ \frac{6 E^3}{k^3} - x^6 \right]^{\frac{1}{6}} \; dx.
\label{BSintex}
\end{equation}
where the limits of integration are defined by the condition $p=0$, which yields $x_{max} = 6^{\frac{1}{6}} \sqrt{\frac{E}{k}}.$

The above integral can be expressed by means of Gamma-functions as follows:
\begin{equation}
2 \sqrt{mk} \; \int_{-x_{max}}^{+ x_{max}} \, \bigg[ \frac{6 E^3}{k^3} - x^6 \bigg]^{\frac{1}{6}} \; dx =
4  \cdot  6^{\frac{1}{3}} \;  \sqrt{mk} \frac{ \big[ \Gamma (\frac{7}{6}) \big]^2}{\Gamma (\frac{4}{3}) } \;  \frac{E}{k},
\label{intexpexp}
\end{equation}
The Bohr-Sommerfeld quantization rule supposes that this integral can take on discrete values $n \hbar$, which leads to the following
discrete values of the energy $E$:
\begin{equation}
E_n = \Bigg[ \frac{\Gamma (\frac{4}{3})}{ 4 \cdot   6^{\frac{1}{3}} \, \left[ \Gamma (\frac{7}{6} ) \right]^2 } \Bigg] \; n \hbar \omega,
\label{E_n}
\end{equation}
with $\omega = \sqrt{k/m}$. Like in the usual harmonic oscillator, the energy levels are proportional to integers, $E_n \sim n \hbar \omega$?
Consequently, the eigenvalues of ${\hat{H}}_{Z_3}$ follow the  rule $\lambda_n \simeq n \omega^3 \hbar^3.$

\section*{Acknowledgements}
The author is greatly indebted to Karol Penson and Michel Dubois-Violette for numerous enlightening discussions
and many useful remarks.

\smallskip

\end{document}